\documentstyle[psfig,amsmath]{ecal97}
\title{Competition in a Fitness Landscape}

\author{Franco Bagnoli$^{(1)}$\thanks{
also INFN and INFM sez. di Firenze; DRECAM-SPEC,
CEA Saclay, 91191 Gif-Sur-Yvette Cedex, France} 
{\rm and}  Michele Bezzi$^{(2)}$\thanks{
INFN, sez. di Bologna}\\
(1) Dipartimento di Matematica Applicata,
Universit\`a di Firenze, \\
via S. Marta,
3 I-50139, Firenze, Italy; e-mail:~bagnoli@dma.unifi.it.\\
(2) Dipartimento di Fisica, 
Universit\`a di Bologna, \\
Via Irnerio 46, I-40126 Bologna, Italy;
e-mail:~bezzi@ing.unifi.it.
\date {}
\vspace{1cm}\\
\begin{minipage}{14cm}
\begin{center} \bf Abstract\end{center}\vspace{1ex}
\normalsize
We present an extension of Eigen's model for quasi-species including
the competition among individuals, proposed as the
simplest mechanism for the formation of new species in a smooth fitness
landscape. 
The evolution equation for the probability distribution of species has the form
of a nonlinear reaction-diffusion equation. 
We are able to obtain analytically an approximation of the 
critical threshold for
the species formation. The comparison with the numerical 
resolution of the original equation is very good.
\end{minipage}
\vspace{1cm}
}

\begin{document}

\maketitle
\thispagestyle{empty}

\section{Introduction}
In this paper we address the problem of formation of species 
in simple ecosystems, possibly mirroring 
some aspects of bacterial and viral
evolution. Our model can be considered as an
extension of Eigen's model~\cite{Eigen71,Eigen:quasispecies}.
With respect to the latter, we introduce 
the interactions among individuals. 

The correspondence of this kind of models with real  biological systems 
is rather schematic: the (haploid) organisms are
only represented by their genetic information (the genotype), 
and we do not consider sexuality nor age structure or polymorphism.
Moreover, a spatial mean field approximation is applied, so that the relevant
dynamical quantity is the distribution of genotypes. This distribution evolves
under the combined effects of selection and mutations. Selection is represented
by the concept of fitness landscape~\cite{Wright32,Peliti95}, 
a function of the genotype that represents the
average fraction of 
survivors per unit of time, and includes the effects of
reproductive efficiency, survival and foraging strategies, predation and
parassitism, etc. In other words, the fitness function is the evolutive
landscape seen by a given individual. 
Only point mutations are considered, and these are  assumed 
to be generated by independent Poisson processes.
The presence of mutations allows the definition of the distance between two
genotypes, 
given by the minimum number of mutations required to connect them.
Assuming only point mutations, the genotypic space is a hypercube,
each direction being spanned by
the possible values of each symbol in the sequence. In the Boolean case, 
which is the one considered here,
each point mutation connects two vertices along the axis corresponding to
the locus where the mutation has taken place.   

In the original work,  Eigen and Schuster~\cite{Eigen:quasispecies} 
showed that a landscape with a single maximum of the fitness allows for a phase
transition from a bell-shaped distribution of the population centered
at the location of the maximum of the fitness function 
(the \emph{master sequence}) to a 
flatter distribution. This \emph{error transition} is triggered by 
the mutation rate.

It has been shown~\cite{Leuthaeusser,Tarazona} that 
Eigen's model is equivalent
to an equilibrium  statistical mechanical model 
of interacting spins, the latter being the elements of the genome. 
 In this way several evolutionary concepts can be mapped to a statistical
mechanics language. In particular, for a static fitness landscape, 
the evolution becomes the process of optimization of an 
``energy'' function (the logarithm of the fitness), 
balanced by the entropy. The genealogy of a particular
genome can be represented as a two-dimensional spin system. 
We refer to the two directions as the time and the
genotypic one, respectively. 
The coupling in the time direction is ferromagnetic and is
given by the mutations. The coupling in the genotypic direction is given by 
the fitness function and  is in general long 
range.
While this mapping is suggestive and allows a precise characterization of
vaguely defined terms, from the point of view of numerical and analytical
treatments of the equations, the original differential equation approach is
more effective. 

Borrowing the 
language of statistical mechanics, 
the single sharp maximum case (the one studied
originally by Eigen and Schuster~\cite{Eigen:quasispecies})
can be defined as a degenerate genotypic space, 
since all individuals but the master sequence have the same fitness,
and we consider it as a particular case of the more general class of
genotypic spaces 
in which the fitness depends only on the genotypic distance 
from the master sequence.  

The degenerate landscape can
be represented as a linear one by introducing the appropriate multiplicity
factor. Using this approach, one implicitly assumes that all degenerate strains
are evenly populated, i.e., that there exist 
high transition rates among these strains.
This assumption has been exploited in the study of the phase
diagram of the single sharp maximum case~\cite{Alves}.   

Finally, assuming a hierarchy for the relevance of mutations, one can have a
pure linear genotypic space. For instance, let us consider Boolean sequences of
length $L$ and assume that $000\dots$ is the master sequence. 
Deleterious mutations 
$0\rightarrow 1$ are assumed to be
non-lethal only if they accumulate at the ends of
the sequence, as (for $L=3$)
\begin{equation}
111 \leftrightarrow 110 \leftrightarrow 100 \leftrightarrow
000 \leftrightarrow 001 \leftrightarrow 011 \leftrightarrow 111,
\label{sequences}
\end{equation}
where the arrows denote the mutations. One can introduce a genotypic index $-3
\le x \le 3$ and rewrite eq.~(\ref{sequences}) as
\begin{equation*}
-3  \leftrightarrow -2 \leftrightarrow -1 \leftrightarrow
0 \leftrightarrow 1 \leftrightarrow 2 \leftrightarrow 3 \equiv -3,
\end{equation*}
i.e., we have a linear genotypic space with periodic boundary conditions.

An hypothetical 
example of such a hierarchical space  is that of a series of genes that code
for enzymes involved in a metabolic pathway. 
A mutation in the first enzyme of the
sequence is more likely lethal, while a mutation  
that lowers the affinity of the last enzyme with its substrate could
be easily retained even if the fitness of the individual is lowered.
This mutation reduce also the specificity of the last enzyme
with its substrate 
(which is the product of the previous metabolic step),
allowing a mutation in the previous enzyme and so on. 

Almost all the works dealt with abstract landscapes (mainly RNA world). 
The difficulty in applying these concepts to real biological systems concerns
the definition of the fitness function, that relates the genotype to the 
phenotype. In particular, the difficulty resides in predicting the stability
or the efficiency of a protein given its sequence of amino acids. One
 can circumvent this difficulty taking into consideration 
only the subclass of all possible mutations
that do not change the protein structure.
One example of an explicit definition of the fitness function is
given by the variation of the reproductive rate of bacteria due to synonymous
mutations and tRNA usage~\cite{BagnoliLio,Giovanna}. This study can be
considered an example of a 
degenerate smooth maximum fitness landscape. Another explicit biological
application concerns the evolution of RNA viruses on HeLa
cultures~\cite{Kessler}. In this case the fitness landscape was assumed to be
linear (without multiplicity). 

While in general the fitness landscape depends  
on the presence of others individuals and changes with time, 
it is  much simpler to study the problem for
a given (static) landscape, that can be thought as an approximation for diluted,
rapidly evolving organisms or self-catalytic molecules (RNA world), while
all other species remain constant. In these
static landscapes, all strains are coupled by the normalization of the
probability distribution, and for small values of the mutation rate 
(a situation fulfilled in the real case) and smooth landscapes, the fittest 
quasi-species always eliminates all
others~\cite{Tarazona}. The global coupling given by the normalization of
probability corresponds to the case of finite population size or the
alternative phases of exponential growth followed by starvation and death. 
For of a rugged static landscape (for instance generated by an Hopfield
Hamiltonian~\cite{Leuthaeusser,Tarazona}),
 the distribution of species can reflect the distribution of the peaks
of fitness. In these cases the interesting question concerns 
the error transition. 

In this work we address the problem of species formation in presence of competition. 
The idea of our approach is the following: we look for a stable probability 
distribution formed by separated quasi-species, and for each of them we compute
the effective fitness landscape due to the competition with individuals of
the same and  all other species. Then, the
parameters of the distribution (position and weight of quasi-species) are
obtained in a self-consistent way. In this way we are able to compute
analytically the threshold for species formation transition in a linear 
landscape. 

We shall deal with coupled differential and finite difference equations, that
can be thought as a mean field approximation of a true microscopic model.
The effect of the finiteness of population, however, should imply a cutoff on 
the tail of the distribution, due to the discreteness of the individuals, and
thus the dependence of evolution on the initial condition (for an application
of the cutoff effect, see Ref.~\cite{Kessler}). 
We do not consider here these effects.

The sketch of this paper is the following: first of all, we formalize the model
in section~\ref{section:the_model}, then we work out the distribution of a
quasi-species near a maximum of the effective fitness landscape in
section~\ref{section:evolution_near_maximum}, and finally we apply the
self-consistency condition in section~\ref{section:speciation}, comparing the 
analytical approximation with the numerical resolution.   
Conclusions and perspectives are drawn in the last section.

\section{The model}
\label{section:the_model}
We describe in detail the approximations that lead to our model. 
An individual is identified by its genome, represented by
an integer index $x$ (no polymorphism nor age structure). 
We study the case of a linear genotypic space (hierarchical relevance of
mutations).

We shall not consider here 
age structure nor the effects of 
polymorphism in the phenotype. 
For the sake of simplicity, we shall deal only with haploid organisms.
Moreover, we do not consider the spatial structure (spatial mean field). 
The experimental setup of reference is that of an bacterial population
that grows in a stirred liquid 
medium, with constant supply of food and removal of
solution, so that the average size of the population is constant. 
Another possible experiment concerns RNA viruses~\cite{Kessler}.

We consider the distribution $p(x,t)$,
that gives the probability of observing the strain $x$ at time $t$ within
the population. We shall denote the whole distribution as $\boldsymbol{p}(t)$.
At each time step we have
\begin{equation}
\sum_{x}p(x,t)=1.
\label{norm}
\end{equation}

Organisms undergo selection, reproduction and mutation. The reproduction and
death rates are represented by a fitness function $A\bigr(x,\boldsymbol{p}(t)\bigl)$, 
that represents the
average fraction of individual of a given strain $x$ 
surviving after a time step in absence of mutation for a given probability
distribution $\boldsymbol{p}(t)$. 

As usual, we consider only point mutations, and we factorize the probability of
multiple mutations (i.e., they are considered independent events). The rate of
mutation per time step of a single element of the genome is $\mu$; 
each point mutation
connects the strain $x$ to $x+1$ or $x-1$.

Since we want to model existing populations, we deal with small
mutation rates. In this limit, only one point mutation can occur at most
during a time step. This is the main difference with previous
works, in which the main goal was to study a mutation-induced phase transition
(error threshold).

With these assumptions, the generic evolution equation (master equation) for the
probability distribution is 
\begin{equation}
\alpha (t)p(x,t+1)=\left(1+\mu \frac{\delta^2}{\delta x^2}\right) 
	A\bigl(x,\boldsymbol{p}(t)\bigl)p(x,t);
	\label{alphap}
\end{equation}
where the discrete second derivative $\delta^2/\delta x^2$ is defined as
\begin{equation*}
\frac{\delta f(x)}{\delta x ^2} = f(x+1) + f(x-1) - 2 f(x),
\end{equation*}
and $\alpha (t)$ maintains the normalization of $\boldsymbol{p}(t)$.
In the following we shall mix freely the 
continuous and discrete formulations of the problem.
 
The numerical resolution of eq.~(\ref{alphap}) shows that a stable asymptotic 
distribution exists for almost all initial conditions.
In the asymptotic limit $t\rightarrow \infty $, $\boldsymbol{p}(t+1)=\boldsymbol{p}(t)
\equiv \boldsymbol{p}(x)$.
Summing over $x$ in eq.~(\ref{alphap}) and using the 
normalization condition, eq.~(\ref{norm}),
we have: 
\begin{equation}
\alpha =\sum_{x}A(x,\boldsymbol{p})p(x)=\overline{A}. 
\end{equation}
The normalization factor $\alpha $ thus corresponds to the average 
fitness. The quantities $A$ and $\alpha $ are defined up to an 
arbitrary constant.

In general the fitness $A$ depends on $x$ and on the probability
distribution $\boldsymbol{p}$. The dependence on $x$ includes the structural
stability of proteins, the efficiency of enzymes, etc. This corresponds to 
the fitness of the individual $x$ if grown in isolation. On the other
hand, the effective fitness seen by an individual depends also on the
composition of the environment, i.e., on $\boldsymbol{p}$. This $\boldsymbol{p}$-dependence
can be further split into two parts:\ the competition with other clones of
the same strain, (intra-strain competition) and that with different strains
(inter-strains competition), disregarding more complex patterns as the group 
structure (colonies).
The intra-strain term has the effect of broadening the curve of a
quasi-species and of lowering its fitness, while the inter-strains part 
can induce the formation of  distinct quasi-species. 

Since $A$ is strictly positive, it can be written as 
\begin{equation*}
A(x,\boldsymbol{p})=\exp \bigl( H(x,\boldsymbol{p})\bigr). 
\end{equation*}
If $A$ is sufficiently smooth (including the dependence on $\boldsymbol{p}$), one can
rewrite eq.~(\ref{alphap}) in the asymptotic limit, using a continuous
approximation for $x$ as 
\begin{equation}
\alpha p=Ap+\mu \frac{\partial ^{2}}{\partial x^{2}}(Ap),
\label{alphap_cont}
\end{equation}
Where we have neglected to indicate
the genotype index $x$ and the explicit dependence on $\boldsymbol{p}$.
Eq.~(\ref{alphap_cont}) has the form of 
 a nonlinear diffusion-reaction equation. Since we want to 
 investigate the phenomenon of species
formation, we look for an asymptotic distribution $\boldsymbol{p}$ formed by a 
superposition of
several non-overlapping bell-shaped curves, where the term non-overlapping 
means almost uncoupled by mutations. Let us number these curves using 
the index $i$, and denote each of them as $p_i(x)$, with
$p(x)=\sum_i p_i(x)$. Each $p_i(x)$ is centered around 
$\overline{x}_i$ and its weight is $\int p_i(x)dx=\gamma _i$, with 
$\sum_i\gamma _i=1$. We further assume that each $p_i(x)$ obeys
 the same asymptotic condition, eq.~(\ref{alphap_cont}) (this is 
 a sufficient but not necessary condition). Defining 
\begin{equation}
\overline{A}_i=\frac{1}{\gamma _i}\int A(x)p_i(x)dx=\alpha,
\end{equation}
we see that in a stable ecosystem all quasi-species have the same average
fitness.

\section{Evolution near a maximum}
\label{section:evolution_near_maximum}

We need the expression of $\boldsymbol{p}$ if a given static fitness $A(x)$
has a smooth, isolated maximum for $x=0$ ({\it smooth maximum} approximation).
Let us assume that 
\begin{equation}
A(x)\simeq A_{0}(1-ax^{2}), 
\label{pot}
\end{equation}
where $A_0 = A(0)$.
Substituting $q=Ap$ in eq. (\ref{alphap_cont}) we have (neglecting to indicate
the genotype index $x$, and using primes to denote differentiation with respect
to it): 
\begin{equation*}
\frac{\alpha }{A}q=q+\mu q''. 
\end{equation*}
Looking for $q=\exp (w)$, %
\begin{equation*}
\frac{\alpha }{A}=1+\mu ({w'}^{2}+w''), 
\end{equation*}
 and approximating $A^{-1}=A_0^{-1}\left( 1+ax^{2}\right) $, we have 
\begin{equation}
\frac{\alpha }{A_{0}}(1+ax^{2})=1+\mu ({w'}^{2}+w''). \label{alphaA0}
\end{equation}
A possible solution is 
\begin{equation*}
w(x)=-\frac{x^{2}}{2\sigma ^{2}}. 
\end{equation*}
Substituting into eq. (\ref{alphaA0}) we finally get 
\begin{equation}
\frac{\alpha }{A_{0}}=\frac{2+a\mu - \sqrt{4a\mu +a^{2}\mu ^{2}}}{2}.
 		\label{smooth}
\end{equation}
Since $\alpha =\overline{A}$, $\alpha /A_{0}$ is less than one we have
 chosen the minus sign. In the limit $a\mu \rightarrow 0$ (small mutation rate
and smooth maximum), we have 
\begin{equation*}
\frac{\alpha }{A_{0}}\simeq 1-\sqrt{a\mu } 
\end{equation*}
and 
\begin{equation}
\sigma ^{2}\simeq \sqrt{\frac{\mu }{a}}. \label{sigma}
\end{equation}

The asymptotic solution is 
\begin{equation*}
p(x)=\gamma \frac{1+ax^{2}}{\sqrt{2\pi }\sigma (1+a\sigma ^{2})}\exp \left(
-\frac{x^{2}}{2\sigma ^{2}}\right),
\end{equation*}
so that $\int p(x)dx=\gamma $. The solution is a bell-shaped curve, its
width $\sigma$ being determined by the combined effects
of the curvature $a$ of maximum and the mutation rate $\mu$.. 
In the next section, we shall apply these results to a quasi-species $i$.
In this case one should substitute $p \rightarrow p_i$, 
$\gamma \rightarrow \gamma_i$ and $x \rightarrow x-\overline{x}_{i}$. 

For completeness, we study also the case of a {\it sharp maximum},  
for which  $A(x)$ varies considerably with $x$. In this case 
the growth rate of less fit strains has a 
large contribution from the mutations of fittest strains, 
while the reverse flow is negligible, thus
\begin{equation*}
p(x-1)A(x-1) \gg p(x)A(x) \gg p(x+1)A(x+1) 
\end{equation*}
neglecting last term, and substituting  $q(x)=A(x)p(x)$ in eq.~(\ref{alphap})
we get: 
\begin{align}
	\frac{\alpha}{A_0}  = 1-2\mu &\qquad \mbox{for $x=0$}\label{map0}\\
 	q(x) =\frac{\mu}{\left(\alpha A(x) 
 	-1+2\mu\right)} q(x-1)&\qquad \mbox{for
 	$x>0$} \label{map}
\end{align}

Near $x=0$, combining eq.~(\ref{map0}), eq.~(\ref{map})and eq.~(\ref{pot})), 
we have
\begin{equation*}
q(x) =\frac{\mu}{(1-2\mu)a x^{2}} q(x-1). 
\end{equation*}

In this approximation the solution is
\begin{equation*}
q(x) = \left(\frac{\mu}{1-2\mu a}\right)^x \frac{1}{(x!)^2},
\end{equation*}
and 
\begin{equation*}
y(x) = A(x)q(x) \simeq \frac{1}{A_0}(1+a x^2)\left(\frac{\mu A_0}{\alpha
a}\right)^x \frac{1%
}{x!^2}. 
\end{equation*}

\begin{figure}[t]
\psfig{figure=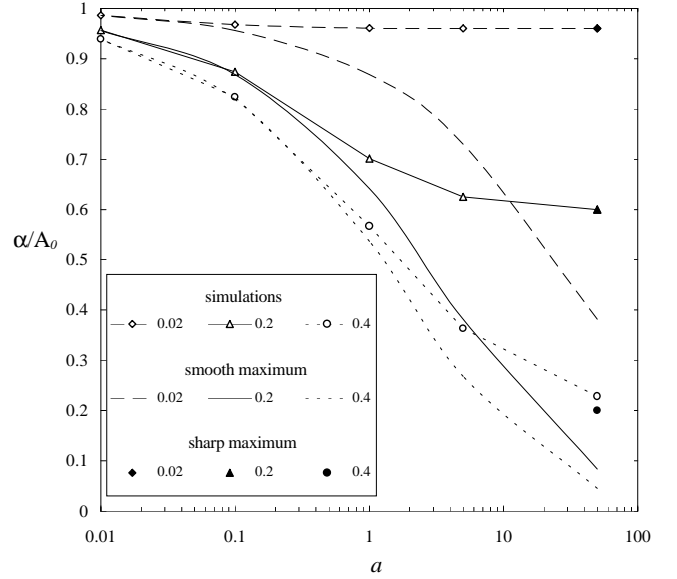,width=9cm}
\caption{\small Average fitness $\alpha/A_{0}$ versus the coefficient $a$,
of the fitness function, eq.~(\ref{pot}),  for some values of the mutation
rate $\mu$. Legend: {\it numerical resolution} corresponds to the numerical solution of 
eq.~(\ref{alphap}),  
{\it smooth maximum} refers to  eq.~(\ref{smooth}) and {\it sharp maximum}
to eq.~(\ref{map0})}
\label{alpha}
\end{figure}

We have checked the validity of these approximations numerically solving  
eq.~(\ref{alphap}); 
the comparisons are shown 
in Figure~(\ref{alpha}).
We observe that the {\it smooth maximum} approximation agrees with the numerics for
for small values of $a$, 
 when $A(x)$ varies slowly with $x$, while the {\it
sharp maximum} approximation  agrees with the  numerical results for large
 values of  $a$, when small variations of $x$ correspond to large
variations of $A(x)$.

\section{Speciation}
\label{section:speciation}

Let us now study the stable quasi-species distribution for a simple 
interacting fitness
landscape. The fitness $A(x,\boldsymbol{p})=\exp (H(x,\boldsymbol{p}))$ is given by 
\begin{equation*}
H(x,\boldsymbol{p})=H_{0}+H_{1}(x)+H_{2}(x,\boldsymbol{p})+\dots 
\end{equation*}
where $H_{0}$ is an arbitrary constant, $H_{1}(x)$ is the static
landscape, i.e., the fitness seen by an individual in isolation (it includes
the interaction with all other slowly varying species) 
and $H_{2}(x,\boldsymbol{p})$ is the interaction
landscape. We examine the case of a single quadratic maximum of $H_{1}$, using
the explicit form: 
\begin{equation*}
H_{1}(x)=b\left( 1-\frac{|x|}{r}-\frac{1}{1+\frac{|x|}{r}}\right),
\end{equation*}
where $r$ gives the amplitude of the quadratic maximum, and $b$ is the
curvature. For $x\rightarrow \infty $, $H_{1}(x)\simeq b(1-|x|/r)$,
 while for $x\rightarrow 0$, $H_{1}(x)\simeq -bx^{2}/r^{2}$.
We have checked numerically that other similar smooth potentials
give the same results of this one. 

We assume that the interactions among 
individuals are always negative (competition) and
decrease exponentially with the distance: 
\begin{equation*}
H_{2}(x,\boldsymbol{p})=-J\int \exp \left( -\frac{\left( x-y\right) ^{2}}{2R^{2}}%
\right) p(y) dy.
\end{equation*}

\begin{figure}[t]
\psfig{figure=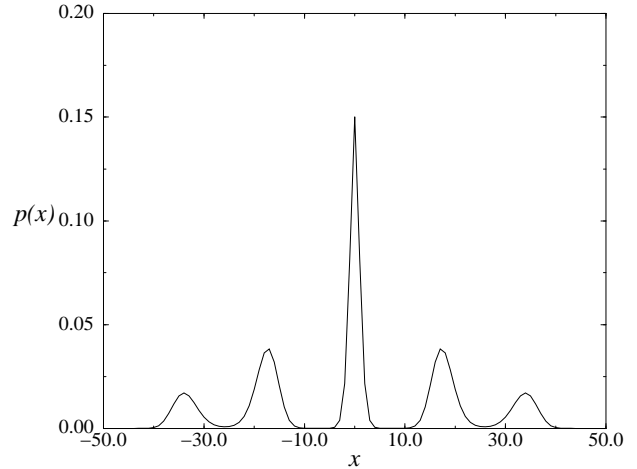,width=9cm,angle=270}
\caption{\small Probability distribution with five quasi-species. Numerical values are 
$\mu=0.01$, $H_0=1.0$, 
$b=0.2$, $J=7.0$, $R=10$ and $r=3$.}
\label{species}
\end{figure}

Numerically solving eq.~(\ref{alphap}) we obtain the asymptotic probability 
distribution showed in Figure~\ref{species}. One can observe the presence of
several non-overlapping quasi-species. 
For $R\rightarrow \infty $, substituting $p(x)=\sum_{i}p_i(x)$, one has 
\begin{equation*}
H_{2}(x,\boldsymbol{p})=-J\sum_{i}\gamma _{i}\exp \left( -\frac{\left( x-\overline{x}%
_{i}\right) ^{2}}{2R^{2}}\right). 
\end{equation*}

The location $\overline{x}_{k}$ of the maximum of the quasi-species
$k$ is given by: 
\begin{equation*}
\left. \frac{dA}{dx}\right| _{\overline{x}_{k}}=0.
\end{equation*}
The species 0 occupies the fittest position $\overline{x}_{0}=0$. For $k\ne
0 $ we have  (using the large $x$ approximation for $H_1$):
\begin{equation*}
-\frac{b}{r}+J\sum_{i}\gamma _{i}\frac{\overline{x}_{k}-\overline{x}_{i}}{%
R^{2}}\exp \left( -\frac{\left( \overline{x}_{k}-\overline{x}_{i}\right)
^{2}}{2R^{2}}\right) =0.
\end{equation*}

We consider now the case of three species, two of which are symmetric with
respect to the dominant one. We have $\overline{x}_{0}=0$, $\overline{x}%
_{1}=-\overline{x}_{2}=\overline{x}$. In the limit $\mu \rightarrow 0$, we
can consider $\alpha =A_{0}(1-\sqrt{a\mu })=A_{0}$, and thus $\alpha =A(0)=A(%
\overline{x})$ (this is a strong approximation which simplifies the
computation), and 
\begin{equation*}
-b+\frac{b\overline{x}}{r}+J\gamma_{1}+\frac{R^{2}b}{r\overline{x}}=J\gamma
_{0}+2\frac{\gamma _{1}}{\gamma _{0}}\frac{R^{2}b}{r\overline{x}}. 
\end{equation*}
Finally, we have the following system 
\begin{align}
	z^{2}-G(\gamma _{0}-\gamma _{1})z+1-2\frac{\gamma _{1}}{\gamma
		_{0}}-z\frac{r}{R} &=0, \\ \label{G}
	\gamma _{0}+2\gamma _{1} &=1, \\ 
	G\gamma _{0}z\exp (-z^{2}/2) &=1,  \label{z}
\end{align}
where $z=\overline{x}/R$ and $G=Jr/Rb$.

The limit of coexistence for the three species is given by $\gamma _{0}=1$
(and thus $\gamma_1=0$).
We compute the critical value $G_c$ of $G$ 
 for the  coexistence of three species, in the limit $r/R\rightarrow
 0$. The first order term $G_c^{(0)}$ is obtained computing $z$
  from eq.~(\ref{G})
\begin{equation*}
z=  \frac{G_c^{(0)}+\sqrt{{G_c^{(0)}}^{2}-4}}{2},
\end{equation*}
and inserting this value into eq.~(\ref{z}). Solving numerically this 
equation, we have $G_c^{(0)}\simeq 2.2160$.
The first correction $G_c^{(1)} (r/R)$ is  obtained from eq.~(\ref{G}), and is
simply $G_c^{(1)}=-r/R$. So finally we have for the critical threshold of
species formation $G_c$
\begin{equation} 
 G_c = 2.216- \frac{r}{R}. \label{Gc}
\end{equation}

\begin{figure}[t]
\psfig{figure=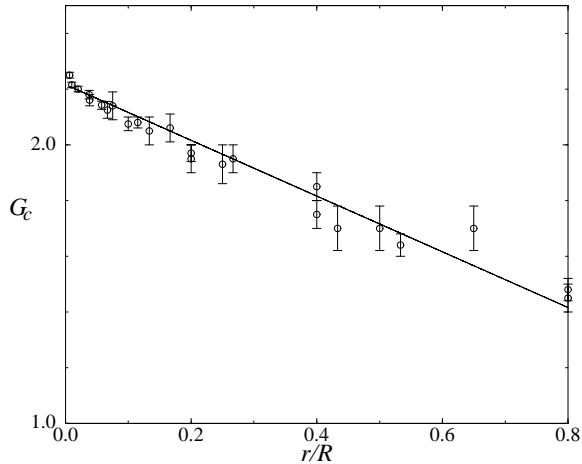,width=9cm,angle=270}
\caption{\small Behavior of $G_c$ versus $r/R$. The continuous line represents the analytical 
approximation, eq.~(\ref{Gc}), the circles are obtained from
 numerical resolution. The error bars represent the maximum error.}
\label{figG}
\end{figure}
 
We have solved numerically eq.~(\ref{alphap}) for 
different values of the parameters,
and we have checked that
the threshold of coexistence of the three species depends only on
$G$. In particular, this threshold does not depends on the mutation rate
$\mu$, at least for $\mu < 0.1$, which is a very high mutation rate for
real organisms. The most important effect of $\mu$ is the broadening of
quasi-species curves, that can eventually merge. 
In the range of parameters used, $G$ depends only 
on ratio $r/R$.  Both these results
are in agreement with the analytical predictions
obtained above. In Figure~\ref{figG} we  compare the numerical and 
analytical results,
plotting the different threshold value $G_c$ as function of $r/R$. 

\section{Discussion and conclusions}
We have studied a simple model for species formation. This model can be considered
an extension of Eigen's one~\cite{Eigen71,Eigen:quasispecies}, with the
inclusion of competition, which if the 
 fundamental ingredient for species formation in smooth landscapes. 
On the other hand, from an individual's point of
view and disregarding complex structures such as  the colonial
 organization, the
more similar the  phenotype the more important the sharing of resources and thus the
competition. Since we assumed a smooth dependence of phenotype on genotype,
we simply modeled the competition $J(x,y)$ between the two strains 
$x$ and $y$ by means of a smooth function of the distance
between two genotypes: $J(x,y) = - J \exp(-(x-y)^2/2R^2)$. 
In this way the strongest competition occurs with other
instances of the same strain, which is reasonable. One can interpret
 our interaction
terms as a cluster expansion of a 
long-range  potential, in which we retained single
and two bodies contributions. From the point of view of population dynamics,
our form of modeling the competition is equivalent to the Verlhust damping
term (logistic equation). 

In a real ecosystem, however, there could be positive contributions to the
interaction term $J$. In particular, it can happen that $J(x,y) > 0$ and
$J(y,x)<0)$ (predation or parassitism), or $J(x,y) > 0$ and $J(y,x)>0)$
(cooperation). An investigation on the origin of complexity in random
ecosystems is in progress. In particular we want to study the effects of 
time fluctuation of fitness (say due to human interaction) on the number of
coexisting species. 

We have studied the effects of competition in a linear (i.e., hierarchic)
genotypic space. 
Our results synthesize in Figure~\ref{figG}. The dependence of the threshold
for the formation of quasi-species obtained analytically from our
approximations reflects very well the numerical results. We also checked that
the latter does not depend on the mutation rate $\mu$, up to $\mu=0.1$. 

\subsection*{Acknowledgements}
We wish to thank G. Guasti, G. Cocho, R. Rechtman, G. Martinez-Mekler and
P.Li\'o for fruitful discussions.
M.B. thanks the Dipartimento di Matematica Applicata ``G. Sansone'' for 
friendly hospitality. Part of this work was done during the workshop
{\it Chaos and Complexity} at ISI-Villa Gualino (Torino, Italy)
under CE contract ERBCHBGCT930295.


\begin{thebibliography}{99}
\bibitem{Eigen71} W. Eigen, \emph{Naturwissenshaften} \textbf{58} 465 (1971).

\bibitem{Eigen:quasispecies} W. Eigen and P. Schuster, 
\emph{Naturwissenshaften} \textbf{64}, 541 (1977).
 
\bibitem{Wright32} S. Wright, \emph{The Roles of Mutation, Inbreeding, 
Crossbreeding, and Selection in Evolution}, \emph{Proc. 6th Int. Cong. 
Genetics, Ithaca}, \textbf{1}, 356 (1932).

\bibitem{Peliti95} L. Peliti, \emph{Fitness Landscapes and evolution} 

{\tt http://xxx.lanl.gov/abs/cond-mat/9505003}

\bibitem{Leuthaeusser} I. Leuth\"ausser, \emph{J. Stat. Phys}
 \textbf{48} 343 (1987).

\bibitem{Tarazona} P. Tarazona, \emph{Phys. Rev. A} \textbf{45} 6038
(1992).
 
\bibitem{Alves} D. Alves and J. F. Fontanari, \emph{Phys. Rev. E} \textbf{54}
4048 (1996). 

\bibitem{BagnoliLio} F. Bagnoli and P. Li\'o, \emph{J. Theor. Biol.}
 \textbf{173} 271 (1995).

\bibitem{Giovanna} F. Bagnoli, G. Guasti, P. Li\'o, \emph{Translation 
optimization in bacteria: statistical models}, in Nonlinear Excitations in
Biomolecules, M. Peyrard, Editor (Les Editions de Phisique-Springer, 1995).

\bibitem{Kessler} L.S. Tsimring, H. Levine and D.A. Kessler, \emph{Phys. Rev.
Lett.} \textbf{76} 4440 (1996).

\end{thebibliography}
\end{document}